\documentclass[prl,twocolumn,showpacs]{revtex4}

\input{epsf}
\usepackage{color}
\newcommand{\bcen}{\begin{center}}
\newcommand{\ecen}{\end{center}}
\newcommand{\btab}{\begin{tabular}}
\newcommand{\etab}{\end{tabular}}
\newcommand{\bdes}{\begin{description}}
\newcommand{\edes}{\end{description}}

\newcommand{\beq}{\begin{equation}}
\newcommand{\eeq}{\end{equation}}
\newcommand{\bea}{\begin{eqnarray}}
\newcommand{\eea}{\end{eqnarray}}
\newcommand{\non}{\nonumber}

\newcommand{\bary}{\begin{array}}
\newcommand{\eary}{\end{array}}
\newcommand{\dou}{\partial}

\newcommand{\prn}[1] {(\ref{#1})}

\newcommand{\fig}[1]{FIG.~\ref{#1}}
\newcommand{\Fig}[1]{FIG.~\ref{#1}}

\usepackage{epsfig}
\newcommand{\D}[1]{\mbox{d}{#1}}

\newcommand{\srr}{\sigma_{rr}}
\newcommand{\stt}{\sigma_{\theta \theta}}
\newcommand{\se}{\sigma_e}
\newcommand{\szz}{\sigma_{zz}}
\newcommand{\err}{\epsilon_{rr}}
\newcommand{\ett}{\epsilon_{\theta \theta}}
\newcommand{\ezz}{\epsilon_{zz}}
\newcommand{\alp}{{\frac{1+3n}{2n}}}

\begin{document}

\bibliographystyle{apsrev}


\title{Dewetting of Glassy Polymer Films} 



\author{Vijay Shenoy$^1$}
\email[]{vbshenoy@iitk.ac.in}
\author{Ashutosh Sharma$^2$}
\email[]{ashutos@iitk.ac.in}
\affiliation{Departments of $^1$Mechanical and $^2$Chemical
Engineering \\ Indian Institute of Technology Kanpur, UP 208 016, India }


\date{\today}

\begin{abstract}
Dynamics and morphology of hole growth in a film of power hardening
viscoplastic solid (yield stress $\sim$ [strain-rate]$^n$) is
investigated. At short-times the growth is exponential and depends on
the initial hole size. At long-times, for $n > \frac{1}{3}$, the
growth is exponential with a different exponent. However, for $n <
\frac{1}{3}$, the hole growth slows; the hole radius approaches an
asymptotic value as $t \rightarrow \infty$. The rim shape is highly
asymmetric, the height of which has a power law dependence on the hole
radius (exponent close to unity for $0.25 < n < 0.4$). The above
results explain recent intriguing experiments of Reiter, {\em
Phys.~Rev.~Lett.}, {\bf 87}, 186101 (2001).
\end{abstract}

\pacs{68.60-p, 61.41.+e, 68.55.-a, 83.50.-v}

\maketitle

Dewetting of largely glassy polymer films below or near their glass
transition temperature is a highly intriguing
phenomenon\cite{Reiter2001} in that the usual mechanisms of shape
change such as viscous fluid flow, surface diffusion,
evaporation/condensation are not dominant\cite{Reiter2001}. Thus, near
glass transition temperature, solid-like plastic yielding and
resulting flow of the glassy polymer appears to be the only mechanism
of dissipation during dewetting.

Indeed, Reiter\cite{Reiter2001} recently observed the growth of holes
in glassy polymer films with features very unlike those
predicted/observed in purely viscous\cite{Redon1991,Oron1997} films.
The key features of the experiments are (i) the radius of the hole
increases with time in an exponential manner (short time behaviour)
(ii) the growth slows down considerably at long times, to the extent
even of nearly stopping\cite{Reiter2002} (iii) the hole grows with a
highly unsymmetric raised rim (iv) the height of the rim is nearly
linearly dependent on the radius of the hole.

In this Letter, we investigate the role of plastic yielding and
resulting flow on the dynamics and morphology of growing holes in
glassy polymer films. Our model is based on two physical premises: (i) the
dewetting hole growth is driven by capillary forces, and (ii) the
polymeric film, near its glass transition temperature, behaves as a
{\em strain-rate hardening viscoplastic solid}\cite{Crist1993}, i.~e.,
the yield stress $\sim$ (effective strain-rate)$^n$, where $n$ is the
strain-rate hardening exponent (henceforth called simply as
``hardening exponent''). The analysis uncovers a rich variety of
regimes of hole growth some of which correspond to the experiments of Reiter\cite{Reiter2001}.

\begin{figure}
\centerline{\input{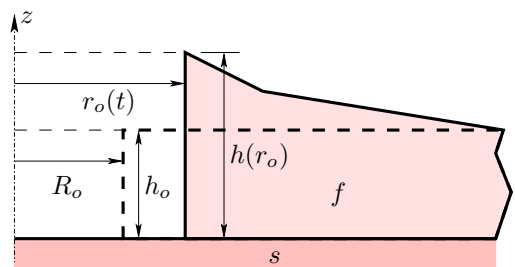}}
\caption{Plane stress model of hole growth. The dashed configuration
is the initial configuration of the film while the solid lines
represent the configuration at time $t$.}
\label{model}
\end{figure}

The model consists of a flat film of initial thickness $h_o$ with a
hole of radius $R_o$(see \fig{model}); we focus on the growth of an
existing hole rather than the mechanism of initial hole formation.  As
the hole grows, the radius increases to $r_o$ and the thickness
becomes inhomogeneous. The aim of the analysis is to obtain the rate
of the hole growth and the profile of the film including the height of
the rim.

Capillary forces are taken to be responsible for hole growth. As the hole
grows there is a net reduction in interfacial energy, the rate of
which is  given by
\bea
\dot{E}_s = 2 \pi \, |S| \, r_o \dot{r}_o \label{Esdot}
\eea
with $|S| = |\gamma_{sa} - \gamma_{sf} - \gamma_{fa}|$ is the spreading
coefficient, where $\gamma_{sf}$ is the interfacial energy between the
film and the substrate, $\gamma_{sa}$ is the surface energy of the
substrate in contact with air and $\gamma_{fa}$ is the surface energy of
the film in contact with air. As the film grows, the height $h(r_o)$
of the rim increases and so does the surface area of the cylindrical
surface of the hole.   The rate of increase in the surface
energy is given by
\bea
\dot{E}_f = 2 \pi \, \gamma_{fa} \, \left(h(r_o) + h'(r_o) r_o \right)
\dot{r}_o. \label{Efdot}
\eea
(We neglect the rate of increase of surface area on the top surface of
the film, and this slightly overestimates the hole growth velocity).
Applying principle of conservation of energy, we get
\bea
\dot{E}_s = \dot{E}_f + D_p \label{dissip}
\eea
where $D_p$ is the rate of plastic dissipation in the deforming film.

The plastic dissipation $D_p$ (see final expression \prn{pdissip}
below) is obtained as follows. 
It is assumed that the film can
slip freely on the substrate, i.~e., the shear stress at the film
substrate interface vanishes -- this assumption is justified by the
fact that the film\cite{Reiter2001} is deposited on a monolayer of PDMS.
Neglecting inertial effects, dynamics provides a relation
between the stress components (a plane stress assumption with a
vanishing component of stress normal to the substrate ($\szz =0$)
is made) 
\bea
\frac{\D \sigma_{rr}}{\D r} + \frac{\sigma_{rr} - \sigma_{\theta
\theta}}{r} = 0. \label{equil}
\eea
where $\sigma_{rr}$ and $\sigma_{\theta \theta}$ are the radial and
circumstantial components of the Cauchy stress tensor.
Nonlinear kinematics of large deformation is treated; the radial
coordinate of the undeformed film is described by $R$ and that of the
deformed film by $r$. The deformation map can be expressed as
$r = r(R,t)$ 
i.~e., a point which was at a radial distance of $R$ at the initial
instance will move to a radial position $r$ at time $t$. In
particular, $r_o(t) = r(R_o,t)$ is the radius of the deformed hole. 
The radial velocity is the time derivative of $r$ defined as
$v = \dou r(R,t)/\dou t = \dot{r}(R,t)$; the radial velocity of 
expansion of the hole is denoted
as $\dot{r_o} = \dot{r}(R_o,t)$.
 The stretching rates $\err$, $\ett$ and $\ezz$ are expressed in terms of the
velocity field as 
\bea
\err = \frac{\dou v}{\dou r}, \;\;\;\; \ett =
\frac{v}{r}, \;\;\;\;
\ezz = -(\err +\ett). \label{kinem}
\eea
 Evidently, the kinematics ensures that the resulting deformation is incompressible.

We neglect elastic strains and treat the deformation as completely
plastic. The polymeric material is taken to be a rate-sensitive
plastic material (viscoplastic material) with power law strain rate
hardening. The effective stress and the effective
plastic strain rate at a material point are related by
\bea
\se =   k \varepsilon^n \label{cons}
\eea
where $\se =\sqrt{\frac{3}{2} s_{ij} s_{ij}}$ with $s_{ij} =
\sigma_{ij} - \frac{1}{3} \sigma_{kk} \delta_{ij}$ is the deviatoric stress tensor, $\varepsilon$ is the
effective plastic strain rate defined as $ \varepsilon =
\sqrt{\frac{2}{3} \epsilon_{ij} \epsilon_{ij}}$, $k$ is a material
constant and $n$ is the hardening exponent. The polymeric material is assumed to obey an associative flow rule\cite{Khan1995},
i.~e., 
\bea
\epsilon_{ij} = \lambda s_{ij} \label{normal}
\eea
where $\lambda$ is a number; the associative rule ensures that the
plastic strain rate is normal to the yield surface. Relations
\prn{equil}, \prn{kinem}, \prn{cons} and \prn{normal} provide a
complete set of equations for investigating hole growth in the
polymeric film.

Expression \prn{normal} can be used to express the nonvanishing
stresses in terms of the stretching rates as
\bea
\srr   =  \frac{2 \err + \ett}{ \lambda}, \;\;\;\; \srr - \stt & = &
\frac{\err - \ett}{ \lambda}. \label{stresses}
\eea
These expressions can be substituted in \prn{cons} to find
$\frac{1}{\lambda} = \frac{2}{3} k \varepsilon^{n-1}$ -- the
stresses are thus expressed completely in terms of the stretching rates.

On substituting the above information into the equilibrium equation
\prn{equil}, we get,
\bea
\frac{\D{}}{\D{r}} \left(\varepsilon^{n-1} \left( 2 \frac{\D{v}}{\D{r}}
+ \frac{v}{r} \right) \right) + \varepsilon^{n-1}\frac{1}{r} \left(\frac{\D{v}}{\D{r}}
-\frac{v}{r}  \right) = 0, ~~~~~ \label{diffeqn}
\eea
where 
\bea
\varepsilon = \sqrt{\frac{2}{3}
\left[\left(\frac{\D{v}}{\D{r}}\right)^2 + \left(\frac{v}{r}\right)^2
+ \left(\frac{\D{v}}{\D{r}} +\frac{v}{r} \right)^2 \right] }. \label{effps}
\eea
The solution of \prn{diffeqn} provides an expression for the radial
velocity  
\bea
v(r) = \dot{r} =  \dot{r_o} \left( \frac{r}{r_o} \right)^{-\frac{1+n}{2n}}. \label{vsol}
\eea
This equation can be used to obtain an expression for the thickness profile of
the film
\bea
h(r,t) = h_o \left( \frac{r}{R} \right)^{\frac{1-n}{2n}}, \;R
= \left(r^\alp - r_o^\alp + R_o^\alp \right)^{\frac{2n}{1+3n}}
. \non \\ \label{heqn}
\eea

To calculate the plastic dissipation, we substitute all the relevant
quantities into the relation 
\bea
D_p = 2 \pi \int_{r_o} ^\infty k \varepsilon^{n+1} \, \, h(r)\, r\, \D{r}
\eea
to obtain
\bea
D_p = 2 \pi \, \alpha
 k h_o  \, r_o^2 \, \left( \frac{\dot{r_o}}{r_o} \right)^{n+1}\, F_n\left(\frac{r_o}{R_o}
\right) \label{pdissip}
\eea
with 
\bea
\alpha & = & 2 \, (3)^{-\frac{1+n}{2}} \,
(1+3n^2)^{\frac{n-1}{2}}\, n^{-n}, \non \\
F_n\left( x \right) & = & \,  _2F_1\left[\mbox{$\frac{1-n}{1+3n}, \frac{1+3n^2}{1+3n};
\frac{2+3n+3n^2}{1+3n}; \left( 1 - x^{-\alp}\right)$}\right], \non
\eea
where $_2F_1$ is the hypergeometric function.

Using \prn{Esdot}, \prn{Efdot} and \prn{pdissip} in \prn{dissip},
the rate of hole growth is obtained as
\bea
\frac{\dot{r_o}}{R_o} = \frac{1}{t_o} \, \left( 1 - \beta \, \left(
\frac{r_o}{R_o}\right)^{\frac{1-3n}{2n}} \right)^{\frac{1}{n}} \,
\left(F_n\left(\frac{r_o}{R_o} \right) \right)^{-\frac{1}{n}}
\frac{r_o}{R_o} \label{rateeqn} \non \\
~
\eea
where $t_o$ is an intrinsic time constant of the film/substrate system
defined as
\bea
t_o = \sqrt[n]{\frac{\alpha k h_o}{|S|}},
\eea
and 
\bea
\beta = \frac{1+n}{2n} \frac{\gamma_{fa}}{|S|} \frac{h_o}{R_o}.
\eea
The rate of hole growth is controlled by two parameters
$\beta$ which is a measure of capillary forces, and $n$ which is the
strain-rate hardening exponent.

The solution to \prn{rateeqn} can be investigated analytically in two
limits, $t \ll t_o$ and $t \gg t_o$. In the short-time limit 
$(t \ll t_o)$, \prn{rateeqn} reduces to
\bea
\frac{\dot{r_o}}{R_o} \approx   \frac{1}{t_o} \, \left( 1 - \beta \right) \,\frac{r_o}{R_o}
\eea
since $F_n(1) = 1$; thus the hole growth is exponential for short
times. It is clear that for the hole to grow the parameter $\beta$ must be
smaller than unity which provides a condition on the initial
radius of the hole that allows for hole growth, i.~e., $\beta < 1$
implies 
\bea
R_o > R_c  = \frac{1+n}{2n}\, \frac{\gamma_{fa}}{|S|} \, h_o 
\eea
where $R_c$ is the critical radius for hole growth $(\beta = R_c/R_o)$. If the
initial radius of the hole is less than $R_c$, the hole does not
grow. As in a fluid film, the analysis for critical hole radius is
most relevant when the film is sufficiently thick to ignore the excess
intermolecular interactions.

\begin{figure}[htb]
\centerline{\epsfxsize=8.0truecm \epsffile{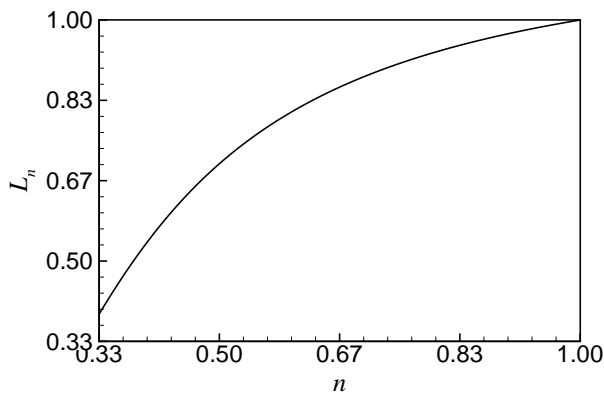}}
\caption{Plot of $L_n$ vs.~$n$.}
\label{Ln}
\end{figure}

The long-time behaviour $(t \gg t_o)$ of the film is investigated in
two different regimes of $n$. First, when $n > \frac{1}{3}$, the rate
of growth \prn{rateeqn} for $ t \gg t_o$ reduces to \bea
\frac{\dot{r_o}}{R_o} \approx \frac{L_n}{t_o} \frac{r_o}{R_o} \eea
$(L_n =\lim_{x \rightarrow \infty} F_n(x)^{-\frac{1}{n}})$ i.~e., the
growth continues to be exponential but with a different exponent
$L_n/t_o$. A plot of $L_n$ as a function of $n$ is shown in
\fig{Ln}. Thus, although the short time exponent depends on the
initial radius $R_o$ of the hole (through $\beta$), the long time
behaviour is independent of the initial size of the hole. Moreover, if
the critical radius is small $(R_c \ll h_o)$, then the long time growth
rate of small unstable holes $(\beta \ll 1)$ slows down compared to
the initial growth rate $L_n < (1-\beta)$.

When $n < \frac{1}{3}$, the long-time behaviour is very
different from the above case. Indeed, the hole radius does not become
indefinitely large, but attains a asymptotic value $r_m$ given by
\bea
r_m = \beta^{-\frac{2n}{1-3n}} R_o
\eea 
A straightforward analysis shows that the approach of $r_o$ towards
$r_m$ is governed by
\bea
\frac{\dot{\overline{\Delta r_o}}}{r_m} \approx -
\frac{1}{t_o}\left(\frac{1-3n}{2n \,F_n(r_m/R_o)}
\right)^{\frac{1}{n}}  \, \left(\frac{\Delta r_o}{r_m} \right)^{\frac{1}{n}}
\eea
where $\Delta r_o = r_m - r_o$. Thus, $r_o$ approaches $r_m$  at a
power law rate governed by the reciprocal of the hardening exponent, 
$1/n$. 

\begin{figure}[htb]
\centerline{\epsfxsize=8.0truecm \epsffile{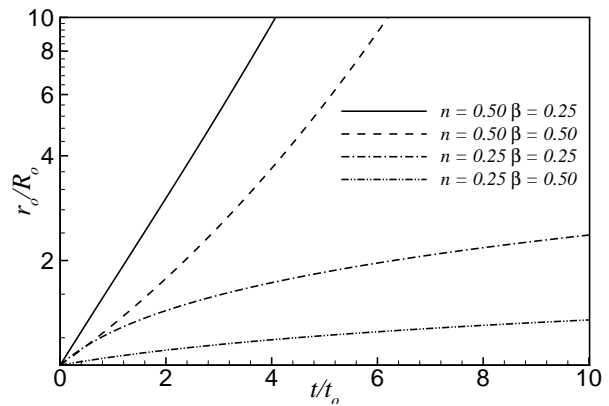}}
\caption{Hole radius as a function of time for various combinations of
$n$ and $\beta$.}
\label{soln}
\end{figure}
The complete solution for \prn{rateeqn} for various combinations  of
$n$ and $\beta$ are shown in \fig{soln}. All the features noted in the
asymptotic analysis are reproduced. It is clear that for a given value
of $\beta$ the rate of growth is larger when $n$ is larger, and for a
given $n$, the rate of growth is larger when $\beta$ is smaller.

The time evolution of the film is also of interest; for $r \gg r_o$,
an asymptotic expression for the increase in the thickness of the film
can be obtained (from \prn{heqn}) as
\bea
\! \! \! \! \! \! \Delta h(r,t) = \frac{(1-n)}{(1+3n)} \frac{( r_o(t)^\alp -
R_o^\alp)}{r^\alp}, \;\; (r \gg r_o)
\eea

\begin{figure}[htb]
\centerline{\epsfxsize=8.0truecm \epsffile{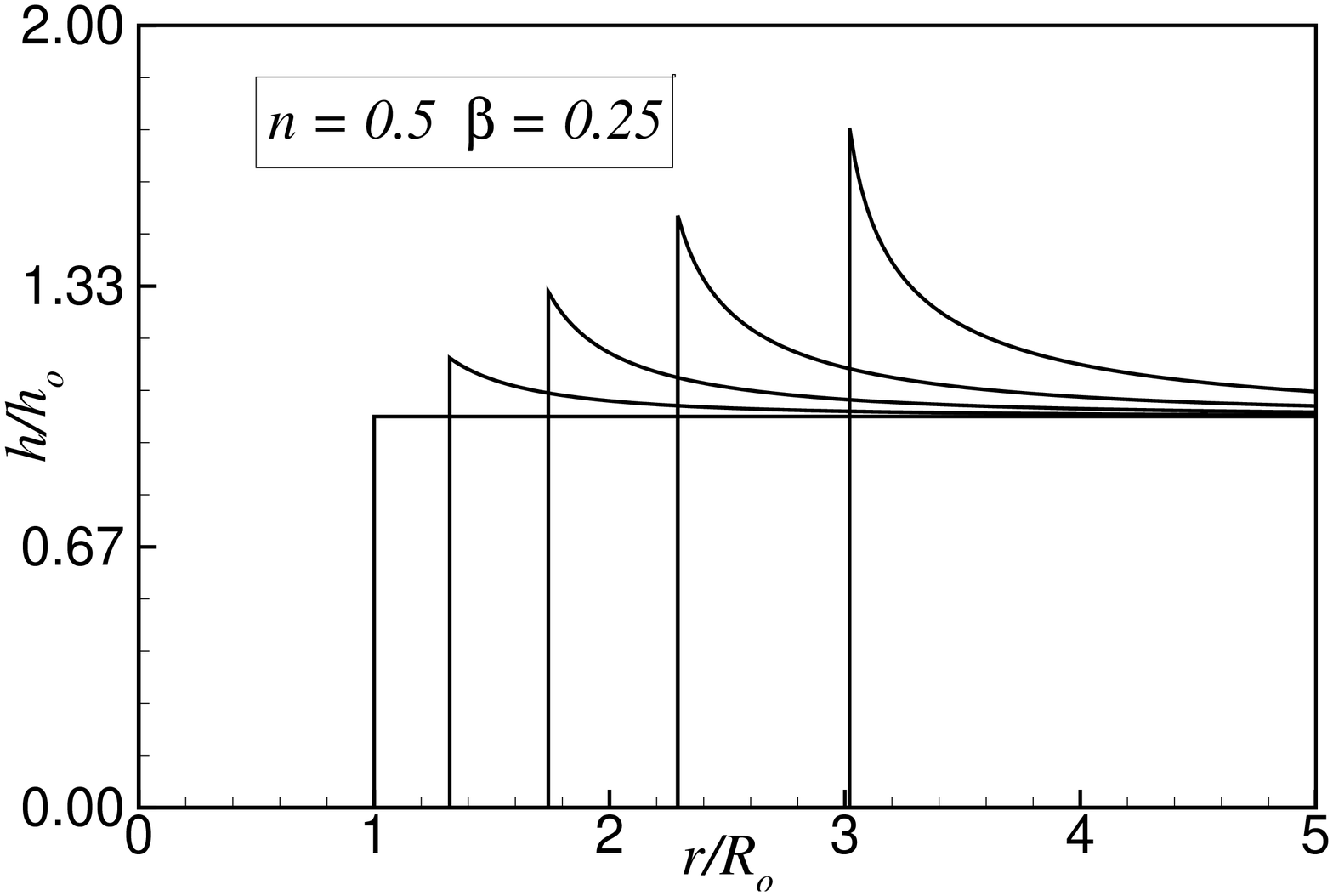}}
\centerline{\epsfxsize=8.0truecm \epsffile{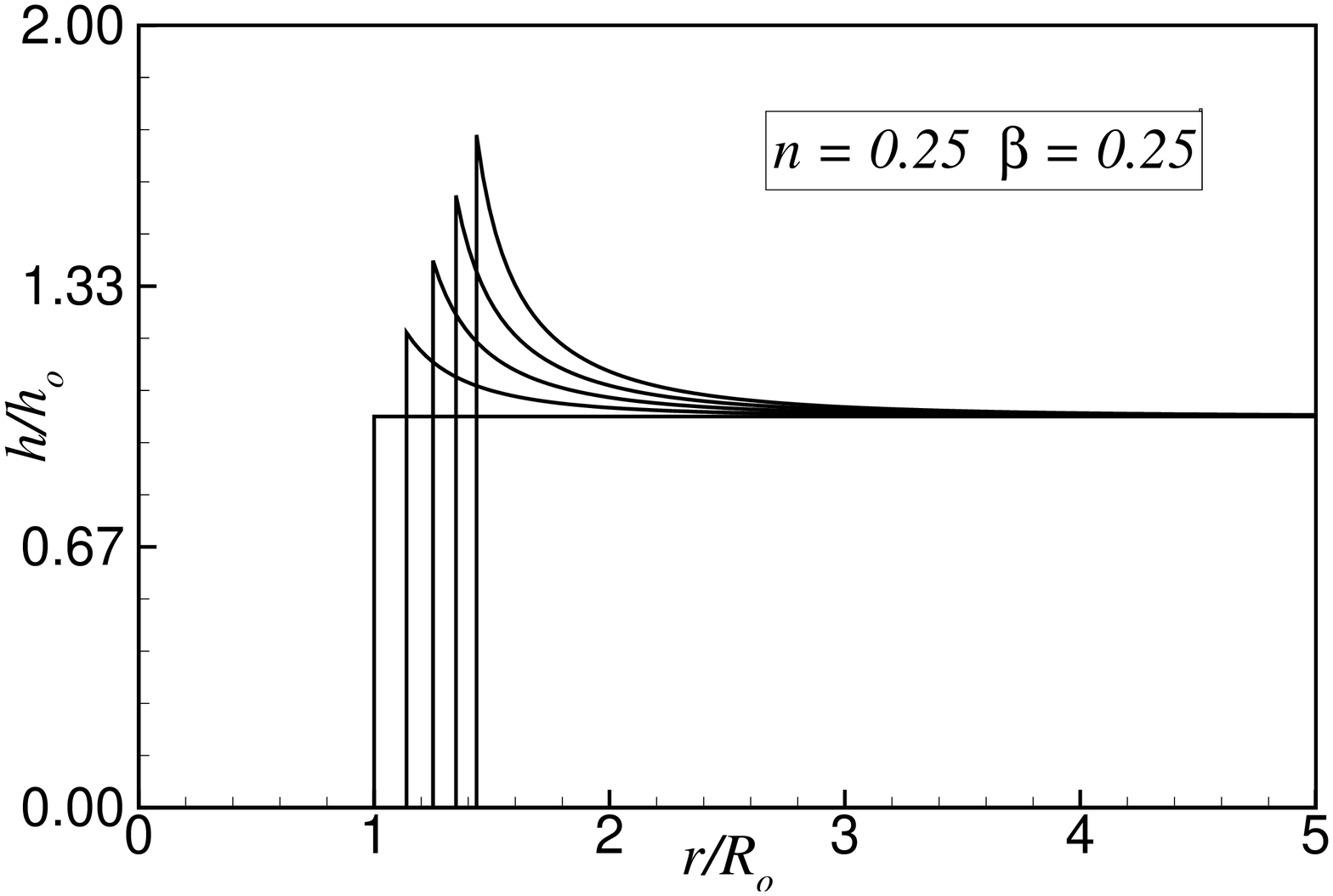}}
\caption{Time evolution of film profiles. The profiles shown are for
$t/t_o=0.0,0.5,1.0,1.5,2.0$.}
\label{tevolve}
\end{figure}
\begin{figure}[htb]
\centerline{\epsfxsize=8.0truecm \epsfbox{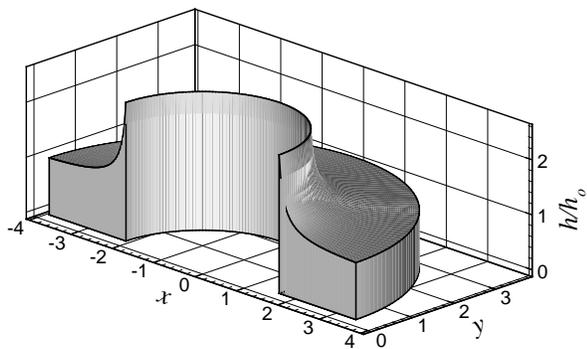}}
\caption{3D depiction of film profile. $n= 0.25, \beta = 0.25,
t/t_o = 5.0$.}
\label{threed}
\end{figure}
\Fig{tevolve} shows  snapshots of  film profiles for two
different values of $n$ for various times. A 3D view of the film
profile is included in \fig{threed} for  visualistion.

Another quantity,  important from the experimental point of view, is the dependence of the height of the rim
on its radius. From \prn{heqn}, it is evident that 
\bea
\frac{h(r_o)}{h_o}  = \left( \frac{r_o}{R_o} \right)^{\frac{1-n}{2n}}
\eea
which is, in general, a power law. However, if $n$ is close to $1/3$,
the exponent is approximately unity. In fact, when $n=0.25$, $h(r_o)
\sim r_o^{1.5}$ and if $n=0.4$, then $h(r_o) \sim r_o^{0.75}$ which
implies that the relationship between the height of the rim and the
radius of the hole is nearly linear in this regime.

The main conclusions of this Letter are summarised as follows: (i) There is a critical value for the initial size of the hole. If
the size of the hole is larger than this value, the hole grows.
(ii) The short time  growth velocity of the hole is
exponential and depends on the initial size of the hole.
(iii) The long time growth velocity of the hole depends on the
strain-rate hardening exponent $n$. When $n > \frac{1}{3}$, the hole
radius grows
exponentially with time with a rate of growth that does not depend on
the initial size of the hole. For $n< \frac{1}{3}$, the hole growth slows down
and the radius of the hole attains an asymptotic value $r_m$ that
depends on the initial size of the hole. 
(iv) The height of the rim is approximately linearly related to the
radius of the hole when $0.25 < n < 0.4$.

The recent observations of Reiter\cite{Reiter2001} on hole growth in
polystyrene films near glass transition temperature can be explained
by the above theoretical results, supporting the premise that the hole
growth in solid polymer films near $T_g$ is governed largely by
yielding and subsequent plastic flow which provides the dominant
dissipative mechanism. The theory also predicts a rich variety of
growth behaviours which may help in the design and interpretation of
new controlled experiments. Our theory for {\em plastically deforming
solids} together with other recent theories of hole growth in {\em
viscoelastic
fluids}\cite{Sapid1996,Brochard-Wyart1997,Jacobs1998,Herminghaus2002}
and {\em shear thinning viscous fluids}\cite{Saulnier2002} provide
complimentary models of dewetting over a wide range of temperatures
from below $T_g$ to well above $T_g$.

Stimulating discussions with G.~Reiter are gratefully
acknowledged. This work was partially supported by Department of
Science and Technology, and Samtel Color Ltd.

\medskip
\hrule
\bibliography{sfb}

\end{document}